\documentclass[twocolumn,9pt]{article} % here we use the article class, rather than elsarticle

\usepackage[square,numbers,sort&compress,comma]{natbib}
\usepackage{amsmath,amssymb}
\usepackage[version=4]{mhchem}
\usepackage{graphicx}
\usepackage{latexsym}
\usepackage{times}          % or better: \usepackage{newtxtext,newtxmath}

\usepackage{enumitem}
\usepackage[pagewise]{lineno}

\usepackage{graphicx}%
\usepackage{amsmath,amssymb,amsfonts}%
\usepackage{amsthm}%
\usepackage{mathrsfs}%
\usepackage[title]{appendix}%
\usepackage{xcolor}%
\usepackage{textcomp}%
\usepackage{manyfoot}%
\usepackage{booktabs}%
\usepackage{algorithm}%
\usepackage{algorithmicx}%
\usepackage{algpseudocode}%
\usepackage{listings}%
\usepackage[colorlinks=true,
            linkcolor=blue,
            citecolor=blue,
            urlcolor=blue]{hyperref}
\usepackage{tabularx}   % Tabellen, die sich automatisch der Breite anpassen
\usepackage{colortbl}
\usepackage{makecell}
\usepackage{pbox}
\usepackage{subcaption}
\usepackage{epsfig}
\usepackage{times}
\usepackage{epsfig}
\usepackage{import}

\usepackage{tikz}
\usepackage{pgfplots}
\pgfplotsset{compat=1.9}
\usetikzlibrary{calc,arrows,arrows.meta,fadings,decorations.pathreplacing,decorations.markings,patterns,shapes.geometric, decorations.pathmorphing}
\tikzset{>=stealth,inner sep=0pt, outer sep=2pt,}
\tikzset{vecteur/.style={->,thick,smooth}}
\usepackage{csvsimple}
\usepackage{xcolor}

\usepackage{siunitx}
\DeclareSIUnit\bar{bar}

\usepackage{multirow}

\topmargin - 12pt % might need to be set to 0pt for some installations
\renewenvironment{abstract}%
              {% - begin definition
               \small% - select font
               {\bfseries \abstractname}% - select font
               \par% - end a paragraph (skip \parsep)
               \vspace{10pt}% - add vertical space
              }% - complete definition

\renewcommand\abstractname{Abstract}

\newcommand{\nomenclature}% - name of command
              [1]% - number of arguments
              {% - begin definition
               \bgroup% - begin a local group
               \flushleft% - turn on flushleft option
               \small\bf% - select font
               #1% - insert title text
               \par% - end a paragraph (skip \parsep)
               \egroup% - terminate local group
              }% - complete definition

\renewcommand{\section}% - name of command
              [1]% - number of arguments
              {% - begin definition
               \bgroup% - begin a local group
               \flushleft% - turn on flushleft option
               \small\bf% - select font
               \refstepcounter{section}% - increment counter
               \arabic{section}. #1% - insert title text
               \par% - end a paragraph (skip \parsep)
               \egroup% - terminate local group
              }% - complete definition

\renewcommand{\subsection}% - name of command
              [1]% - number of arguments
              {% - begin definition
               \bgroup% - begin a local group
               \flushleft% - turn on flushleft option
               \small\em% - select font
               \refstepcounter{subsection}% - increment counter
               \arabic{section}.% - insert title text
               \arabic{subsection}. #1% - insert title text
               \par% - end a paragraph (skip \parsep)
               \egroup% - terminate local group
              }% - complete definition

\renewcommand{\subsubsection}% - name of command
              [1]% - number of arguments
              {% - begin definition
               \bgroup% - begin a local group
               \flushleft% - turn on flushleft option
               \small\em% - select font
               \refstepcounter{subsubsection}% - increment counter
               \arabic{section}.% - insert title text
               \arabic{subsection}.% - insert title text
               \arabic{subsubsection}. #1% - insert title text
               \par% - end a paragraph (skip \parsep)
               \egroup% - terminate local group
              }% - complete definition

  \newcommand{\acknowledgement}% - name of command
              [1]% - number of arguments
              {% - begin definition
               \bgroup% - begin a local group
               \flushleft% - turn on flushleft option
               \small\bf% - select font
               #1% - insert title text
               \par% - end a paragraph (skip \parsep)
               \egroup% - terminate local group
              }% - complete definition

  \newcommand{\sectionbib}% - name of command
              [1]% - number of arguments
              {% - begin definition
               \bgroup% - begin a local group
               \flushleft% - turn on flushleft option
               \small\bf% - select font
               #1% - insert title text
               \par% - end a paragraph (skip \parsep)
               \egroup% - terminate local group
              }% - complete definition

\begin{document}

% -------------------------------------------------------------------- %
% -------------------------------------------------------------------- %
% -------------------------------------------------------------------- %

% -------------------------------------------------------------------- %

\small
\baselineskip 10pt

% -------------------------------------------------------------------- %
% -------------------------------------------------------------------- %
% -------------------------------------------------------------------- %
\setcounter{page}{1}
% -------------------------------------------------------------------- %
\title{\LARGE \bf A Novel Approach for Direct Measurement of the Stretch Factor in Laminar Premixed Hydrogen–Air Flames Affected by Thermodiffusive Instabilities}

\author{{\large M. Marburger$^{a,*,\dagger}$, C. Möller$^{a,\dagger}$, A.R.W. Macfarlane$^{a}$,}\\
        {\large M. Schneider$^{b}$, B. Traut$^{b}$, C. Hasse$^{b}$, A. Gruber$^{c,d}$, A. Dreizler$^{a}$}\\[0pt]
        {\footnotesize \em $^a$Technical University of Darmstadt, Department of Mechanical Engineering, Reactive Flows and Diagnostics}\\[-5pt]
        {\footnotesize \em Otto-Berndt-Stra{\ss}e 3, 64287 Darmstadt, Germany}\\[-5pt]
        {\footnotesize \em $^b$Technical University of Darmstadt, Department of Mechanical Engineering,}\\[-5pt]
        {\footnotesize \em Simulation of Reactive Thermo-Fluid Systems, Otto-Berndt-Stra{\ss}e 2, 64287 Darmstadt, Germany}\\[-5pt]
        {\footnotesize \em $^c$SINTEF Energy Research, Thermal Energy Department, Trondheim N-7465, Norway}\\[-5pt]
        {\footnotesize \em $^d$Norwegian University of Science and Technology, Department of Energy and Process Engineering,}\\[-5pt]
        {\footnotesize \em N-7491 Trondheim, Norway}\\[-20pt]
}
\date{}  %%% Leave as is, do not add date;

% -------------------------------------------------------------------- %
% -------------------------------------------------------------------- %
% -------------------------------------------------------------------- %
\twocolumn[\begin{@twocolumnfalse}
\maketitle
\rule{\textwidth}{0.5pt}
\vspace{-5pt}

\begin{abstract} % 100 to 300 words.
The present study introduces a novel experimental configuration employing optical OH-PLIF imaging to directly determine the stretch factor ($I_0$) in laminar premixed hydrogen flames as they transition from a quasi-stable to a thermodiffusively unstable regime.
The setup consists of a rod-anchored V-shaped flame stabilised in a laminar flow of premixed reactants.
In the near field of the anchoring rod, the mildly strained flame remains quasi-stable, characterised by a smooth surface and a well-defined inclination angle ($\theta_{\mathrm{s}}$) relative to the main flow.
This region defines the first, \textit{stable-branch} of the V-flame, with a corresponding burning rate $S_{\mathrm{s}}$.
Further downstream, the flame abruptly transitions into a distinct regime dominated by pronounced thermodiffusive (TD) instabilities, as evidenced by cellular structures and a strongly wrinkled flame surface.
The distance between this transition point and the anchor decreases with increasing equivalence ratio.
This second \textit{TD-unstable branch} is characterised by a marked increase in the mean flame-surface angle ($\theta_{\mathrm{u}}$) relative to the flow direction, allowing direct evaluation of the increase in global flame speed, $S_{\mathrm{u}}/S_{\mathrm{s}}$, between the stable and TD-unstable branches.
Notably, it is assumed that this ratio represents the normalised flame consumption speed $S_{\mathrm{c}}/S_{\mathrm{L}}$.
Determination of $I_0$ additionally requires the ratio of increase in surface-area due to the thermodiffusive instabilities (TDI). 
Three complementary methods are employed to evaluate the flame-surface area of the TD-unstable branches ($A$), comparing it to a smooth reference area ($A_0$), yielding consistent trends in $A/A_0$ across the range of equivalence ratios studied.
The resulting $I_0$ estimates, while subject to uncertainties primarily in $A$, decrease monotonically with increasing equivalence ratio, from approximately $1.1$--$1.3$ at $\phi = 0.35$ to $0.8$--$0.9$ at $\phi = 0.4$, which is consistent with theoretical predictions.
Additional numerical simulations in a reduced two-dimensional representation of the experimental configuration show the same transition behaviour and yield qualitatively consistent results.
\end{abstract}

\vspace{10pt}

{\bf Novelty and significance statement}
\vspace{10pt}

%< 150 words!
This work introduces a novel method for directly measuring the stretch factor $I_0$ from OH-PLIF imaging of rod-anchored lean premixed laminar hydrogen-air V-flames.
The method exploits the abrupt transition from a quasi-stable regime with a nearly flat flame surface to a thermodiffusively unstable regime with highly wrinkled flame surfaces.
In contrast to jet burners or spherical expanding flames, the proposed configuration enables the onset of thermodiffusive instabilities to be investigated largely independently of configuration-induced effects on the flame dynamics, such as imposed curvature and flame-branch interactions.
The well-characterised laminar flow field allows the transition to be reproduced systematically with minimal external perturbations and clearly identified over a relevant range of fuel-lean operating conditions.
Complementary numerical simulations in a reduced two-dimensional representation of the experimental configuration reproduce the same qualitative transition behaviour and support the experimental findings.
Together, the experimental and numerical results provide a valuable basis for theory development and validation of numerical models for flames prone to thermodiffusive instabilities.

\vspace{5pt}
\parbox{1.0\textwidth}{\footnotesize {\em Keywords:}
Premixed Hydrogen-Air Flame;
Thermodiffusive Instability (TDI);
Stretch factor $I_0$}
\rule{\textwidth}{0.5pt}
*Corresponding author.
$^\dagger$ Authors with equal contribution.
\vspace{5pt}
\end{@twocolumnfalse}] 

% \linenumbers
\section{Introduction\label{sec:introduction}} \addvspace{10pt}
Hydrogen (\ce{H2}) is a key energy carrier for decarbonisation~\cite{Dreizler2021}.
Lean premixed operation helps mitigate \ce{NO_{x}} emissions in industrial applications.
However, the microscopic structure of sub-unity Lewis number flames is strongly affected by preferential and differential diffusion~\cite{Pitsch2024}, making fuel-lean \ce{H2} flames susceptible to intrinsic thermodiffusive instabilities (TDIs) and thus inherently multidimensional.
TDIs enhance the local flame speed and flame surface area (due to the formation of cellular and finger-like structures) \cite{Altantzis2012}, lead to superadiabatic temperatures, may influence flame flashback \cite{Porath2025}, and have recently also been shown to significantly affect flame--wall interaction (FWI) \cite{Schneider2025a, Schneider2025b, Schneider2025c, Marburger2026}.
The cellular structures characterising TDIs have been observed for decades~\cite{Zeldovich1944, Markstein1949, Sivashinsky1977, Matalon2003}, and the underlying physical mechanisms are detailed, for example, by Williams~\cite{Williams1985}.
Under turbulent conditions, turbulent eddies interact nonlinearly with preferential and differential diffusion, thereby further enhancing the thermodiffusive (TD) response up to a certain limit~\cite{Aspden2011, Aspden2015, Aspden2019, Aspden2024, Berger2022c, Berger2024, Shi2024b}.
TDIs have also been investigated in spherically expanding flames~\cite{Bradley1994,Yang2018}, Bunsen burners~\cite{Shi2024a}, and jet flames~\cite{Li2025, Shi2024b}, demonstrating the important roles of turbulence and flame geometry.
Moreover, TDIs have been identified in practical combustion systems, such as internal combustion engines, in both experiments~\cite{Ye2025} and numerical studies~\cite{Traut2026}, emphasizing their significance for real-world operation.

In general, the combined effects of turbulence, TDIs, preferential and differential diffusion increase the global flame consumption speed through both an enlargement of the flame surface area and an enhancement of the local flame speed, the latter commonly quantified by the normalised mean local flame speed (stretch factor) $I_0$~\cite{Bray1991}.
While multidimensional simulations of freely propagating flames can capture intrinsic instabilities~\cite{Kadowaki2005b, Altantzis2012, Berger2022b, Howarth2022, Howarth2023, Nicolai2025} and allow $I_0$ to be extracted with high accuracy, their computational cost, particularly for fully resolved three-dimensional cases, limits their use in large parametric studies.
Nevertheless, empirical models for $I_0$ have been developed from two- and three-dimensional simulations of TDIs~\cite{Howarth2022, Howarth2023, Rieth2023}.

%Experimental measurements
Experimental measurements of $I_0$ in hydrogen premixed flames remain scarce. A recent study by Chaib et al.~\cite{Chaib2024} reported measurements of $I_0$ in weakly turbulent hydrogen-enriched methane/air flames based on OH-PLIF, utilising OH-intensity gradient magnitudes as markers of TDIs.

%In this study, a V-flame configuration, previously used for hydrocarbon flames~\cite{Bell2005, Kosaka2018}, is adapted to study laminar, TD-unstable hydrogen-air flames at fuel-lean conditions using OH-PLIF.
In this study, a V-flame configuration previously used to investigate FWI in methane, DME and hydrogen flames ~\cite{Kosaka2018, Zentgraf2022, Marburger2026} is adapted to study laminar, TD-unstable hydrogen-air flames at fuel-lean conditions using OH-PLIF, without wall interaction.

The diverging branches prevent tip interactions, allowing TDIs to develop freely.
The \ce{H2} flame exhibits two regimes: an upstream smooth front (\textit{stable-branch}) and a downstream wrinkled front (\textit{TD-unstable branch}) with abruptly increased flame angle and local propagation speed at TDI onset.
It is proposed that the stretch factor $I_0$ can be determined from combined measurements of flame angle and flame surface area, without requiring direct measurement of the fuel consumption speed.
Experiments across fuel-lean equivalence ratios use multiple methods to estimate the flame surface area, highlighting the configuration’s utility for both qualitative and quantitative TDI characterisation.
Additionally, numerical simulations in a reduced two-dimensional representation of the experimental configuration are performed to qualitatively confirm the mechanisms, methodology, and trends observed in the experiments.

% I would really like, if we could formulate research questions / objectives and we should probably highlight the novelty more
First, the experimental setup and the corresponding data-processing methods are described in Sec.~\ref{sec:experimental_setup}, followed by the numerical setup and the associated evaluation methods in Sec.~\ref{sec:numerical_setup}.
Based on the experimental and numerical analyses, the results are discussed in Sec.~\ref{sec:results}.
Finally, the conclusions are presented in Sec.~\ref{sec:conclusion}.

\section{Experimental Setup and Evaluation\label{sec:experimental_setup}} \addvspace{10pt}
\begin{figure}[htb]
    \centering
    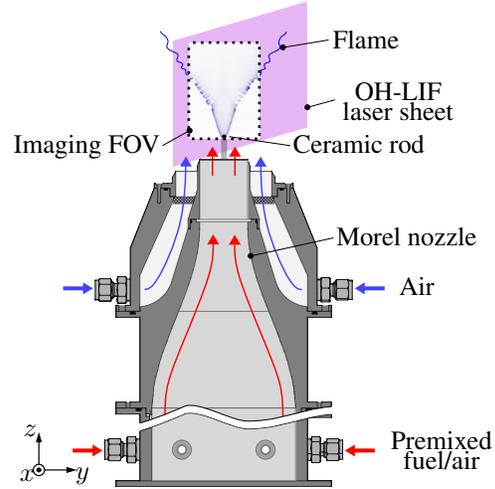
    \captionsetup{aboveskip=4pt, belowskip=-10pt}
    \caption{Cross section of the test rig with imaging FOV.}
    \label{fig:FWISketch}
\end{figure}

\begin{figure*}[ht]
    \centering
    \def\svgwidth{\textwidth}
    %% Creator: Inkscape 1.4.3 (0d15f75, 2025-12-25), www.inkscape.org
%% PDF/EPS/PS + LaTeX output extension by Johan Engelen, 2010
%% Accompanies image file 'AllOPs_Onset_v5.pdf' (pdf, eps, ps)
%%
%% To include the image in your LaTeX document, write
%%   \input{<filename>.pdf_tex}
%%  instead of
%%   \includegraphics{<filename>.pdf}
%% To scale the image, write
%%   \def\svgwidth{<desired width>}
%%   \input{<filename>.pdf_tex}
%%  instead of
%%   \includegraphics[width=<desired width>]{<filename>.pdf}
%%
%% Images with a different path to the parent latex file can
%% be accessed with the `import' package (which may need to be
%% installed) using
%%   \usepackage{import}
%% in the preamble, and then including the image with
%%   \import{<path to file>}{<filename>.pdf_tex}
%% Alternatively, one can specify
%%   \graphicspath{{<path to file>/}}
%% 
%% For more information, please see info/svg-inkscape on CTAN:
%%   http://tug.ctan.org/tex-archive/info/svg-inkscape
%%
\begingroup%
  \makeatletter%
  \providecommand\color[2][]{%
    \errmessage{(Inkscape) Color is used for the text in Inkscape, but the package 'color.sty' is not loaded}%
    \renewcommand\color[2][]{}%
  }%
  \providecommand\transparent[1]{%
    \errmessage{(Inkscape) Transparency is used (non-zero) for the text in Inkscape, but the package 'transparent.sty' is not loaded}%
    \renewcommand\transparent[1]{}%
  }%
  \providecommand\rotatebox[2]{#2}%
  \newcommand*\fsize{\dimexpr\f@size pt\relax}%
  \newcommand*\lineheight[1]{\fontsize{\fsize}{#1\fsize}\selectfont}%
  \ifx\svgwidth\undefined%
    \setlength{\unitlength}{472.32002318bp}%
    \ifx\svgscale\undefined%
      \relax%
    \else%
      \setlength{\unitlength}{\unitlength * \real{\svgscale}}%
    \fi%
  \else%
    \setlength{\unitlength}{\svgwidth}%
  \fi%
  \global\let\svgwidth\undefined%
  \global\let\svgscale\undefined%
  \makeatother%
  \begin{picture}(1,0.26346543)%
    \lineheight{1}%
    \setlength\tabcolsep{0pt}%
    \put(0,0){\includegraphics[width=\unitlength,page=1]{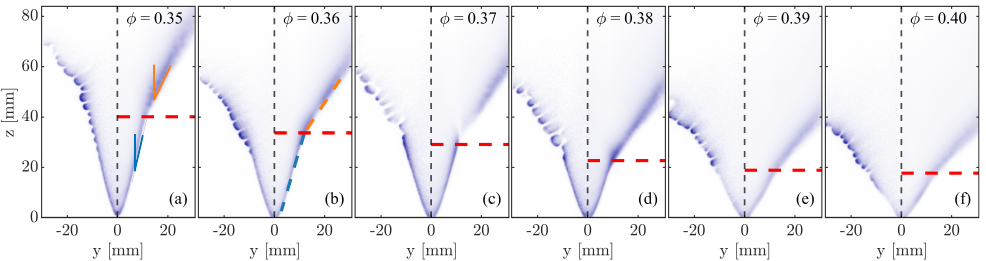}}%
    \put(0.15629423,0.10055245){\color[rgb]{0.12156863,0.46666667,0.70588235}\makebox(0,0)[t]{\lineheight{0.30000001}\smash{\begin{tabular}[t]{c}$\theta_\text{s}$\end{tabular}}}}%
    \put(0.18018171,0.17181071){\color[rgb]{1,0.49803922,0.05490196}\makebox(0,0)[t]{\lineheight{0.30000001}\smash{\begin{tabular}[t]{c}$\theta_\text{u}$\end{tabular}}}}%
    \put(0,0){\includegraphics[width=\unitlength,page=2]{AllOPs_Onset_v5.pdf}}%
  \end{picture}%
\endgroup%

    \captionsetup{aboveskip=4pt, belowskip=-10pt}
    \caption{Selected single shots (left panels) and averages (right panels) for all equivalence ratios.
    The horizontal dashed red lines indicate the mean onset location of TDIs. An example of the calculated fit segments is given for the equivalence ratio of $\phi=0.36$. The fit of the stable section is indicated by the blue dashed line, the fit of the unstable section by the orange dashed line.}
    \label{fig:AllOPs}
\end{figure*}

Measurements are conducted using the test rig shown in Fig.~\ref{fig:FWISketch}, where a V-shaped premixed H\textsubscript{2}/air flame is stabilised on a $\SI{1}{\milli\metre}$-diameter ceramic rod as a flame anchor. 
The flame-stabilising rod is positioned $\SI{18}{\milli\meter}$ downstream of the nozzle and centred length-wise in the x-direction. Since the rod is not temperature-controlled, its temperature increases due to heat transfer from the flame. However, during measurements steady-state conditions were ensured.
This effect is taken into account in the simulations (Sec.~\ref{sec:numerical_setup}).

The burner is operated at ambient conditions ($\SI{1}{\bar}$, $\SI{293.15}{\kelvin}$) with six equivalence ratios from $0.35$ to $0.4$ at a constant bulk velocity $u_{bulk}$ of $\SI{2.28}{\frac{\meter}{\second}}$.
Flow issues from a $\SI{40}{\milli\metre} \times \SI{40}{\milli\metre}$ Morel nozzle designed for a top-hat velocity profile.
A premixed H\textsubscript{2}/air mixture is supplied via an inlet plenum $\SI{100}{\milli\metre}$ upstream, followed by homogeniser grids and a honeycomb section to ensure uniform flow.
A co-flow of air surrounds the flame to minimise disturbances with a constant velocity of $\SI{0.9}{\metre\per\second}$ for all operating conditions. All mass flows are controlled with calibrated Bronkhorst controllers (accuracy error below $1.1\%$ in the operated range).
Flame fronts are visualised using OH planar laser-induced fluorescence (OH-PLIF) excited at the Q\textsubscript{1}(6) line of the $\mathrm{A}^2\Sigma^+ - \mathrm{X}^2\Pi(1,0)$ transition, and imaged onto an intensified CCD camera.
The field of view is $\sim\SI{60}{\milli\meter} \times \SI{80}{\milli\meter}$, corresponding to $\sim\SI{60}{\micro\metre/px}$, with a laser repetition rate of $\SI{10}{\hertz}$.
The optical resolution is determined to be $\sim\SI{300}{\micro\meter}$ using a Siemens star and determined based on the Rayleigh criterion.

\subsection{Experimental Image Processing and Data Evaluation \label{sec:processing}} \addvspace{10pt}

Figure~\ref{fig:AllOPs} presents OH-PLIF measurements for all investigated operating conditions.
For each case, a representative instantaneous snapshot is shown on the left, while the temporally averaged OH field is displayed on the right. 
The averaged images serve as the basis for the quantitative analysis described below.

For each operating condition, 200 single-shot images were averaged. The two branches of the V-shaped flame were analysed separately by restricting the evaluation to either the left or right branch. Both branches were processed independently and the resulting values were averaged, effectively doubling the number of samples.

To avoid interference from the anchoring structure, a region extending approx. $\SI{3}{\milli\metre}$ downstream of the stabilising rod was excluded. Spatial gradients of the averaged images were computed in the horizontal and vertical directions, and the gradient magnitude was subsequently determined using both. The flame front was extracted row-wise by identifying the position of maximum gradient magnitude, with the search restricted to horizontal gradients corresponding to an increase in OH signal.

The extracted flame fronts were then used to determine the onset of TDI cells. The mean flame front was approximated by two linear segments representing the upstream (TD-stable) and downstream (TD-unstable) regions. Both fits were constrained to intersect within the evaluated domain, and the intersection point was defined as the TDI onset location.

The boundary between the TD-stable and TD-unstable regions was determined by systematically shifting a candidate split point along the extracted flame front.
For each candidate position, the flame front points were separated into two subsets ($N_1$ and $N_2$).
Independent linear regressions were then performed on each subset, resulting in two linear polynomials ($p_1$ and $p_2$).
The overall fit quality was evaluated using a global root-mean-square error (RMSE) calculated across all flame front points.
The RMSE quantifies the deviation between the measured flame front positions and those predicted by the two linear fits:

\begin{equation}
\mathrm{RMSE} =
\sqrt{\frac{1}{N_1+N_2}
\left(
\sum_{i=1}^{N_1} r_{1,i}^2 +
\sum_{j=1}^{N_2} r_{2,j}^2
\right)}
\end{equation}

Here, the residuals are defined as
$r_{1,i} = x_{1,i} - p_1(y_{1,i})$ for $i = 1,\dots,N_1$ and
$r_{2,j} = x_{2,j} - p_2(y_{2,j})$ for $j = 1,\dots,N_2$, representing the deviation between the measured flame front positions and those predicted by the respective linear fits.

The split position producing the lowest global RMSE was selected as the optimal separation between the TD-stable and TD-unstable flame regions.
The intersection of the corresponding linear fits defines the detected TDI onset location (horizontal red dashed lines in Fig.~\ref{fig:AllOPs}).
%This procedure was applied independently to both branches of the V-flame, and the mean of the two resulting intersection points was used to indicate the TDI onset (red dashed lines in Fig.~\ref{fig:AllOPs}).

The final linear fits additionally provide the inclination angles of the upstream and downstream flame segments, $\theta_{\mathrm{s}}$ and $\theta_{\mathrm{u}}$, respectively, as illustrated in Fig.~\ref{fig:AllOPs}.
From these angles, the ratio of burning velocities in the unstable and stable branch ($S_{\mathrm{u}}/S_{\mathrm{s}}$) can be obtained geometrically.
%The same analysis procedure was also applied to the temporally averaged numerical data.

To quantify the flame-surface area increase downstream of the onset, the same analysis is applied to single-shot images. Using the previously described detection procedure, the downstream TD-unstable branch is isolated and rotated according to the extracted inclination angle of the unstable section such that the branch becomes horizontal. The ratio of wrinkled to unwrinkled flame contour lengths, ${A}/{A_0}$, is then used as a measure of flame surface area enhancement. For the same isolated region, $A_0$ represents the idealised flame surface without wrinkles.

Three flame front detection strategies were used to assess robustness. The first, ``f-canny'', is a gradient-based detector operating directly on the isolated area of interest (AoI). A Canny-type edge detector (introduced in~\cite{canny2009computational}) is applied after preprocessing, and the resulting edge map is filtered using gradient-magnitude masking, gradient-direction masking and connected-component size filtering to suppress weak or spurious structures.

The second method, ``f-otsu'', is based on Otsu thresholding (introduced in~\cite{otsu1979threshold}) and perimeter extraction of segmented intensity regions. For this the images were preprocessed, by adaptive histogram equalisation, gamma correction and gaussian smoothing. The processed AoI is binarised by the minimum of of the Otsu threshold and the median post-flame intensity. Small binary regions are removed, holes are filled, and the perimeter of the dominant flame region is extracted. The resulting perimeter is then filtered by edge orientation, connected-component size, and an overlap criterion removing edge segments of which more than half is above another segment in the AoI.

The third method ``f-otsuCanny'', similar to that proposed in~\cite{Chaib2024}, combines threshold-based and gradient-based detection. An Otsu-derived flame-region perimeter is first constructed using the same preprocessing, segmentation and filtering procedure as in ``f-otsu'' without the directional filtering. A spatial band is then defined around this perimeter. Independently, a Canny edge detector is applied to the raw AoI, and the final flame front is obtained by restricting the Canny result to this Otsu-derived band followed by small-component filtering.

For all three methods neighbouring flame-contour segments are then connected with a straight line from endpoint to endpoint, up to gaps of $\approx\SI{2.5}{\milli\meter}$. More details on the processing methods are presented in the supplementary material.

\section{Numerical Setup and Evaluation\label{sec:numerical_setup}} \addvspace{10pt}
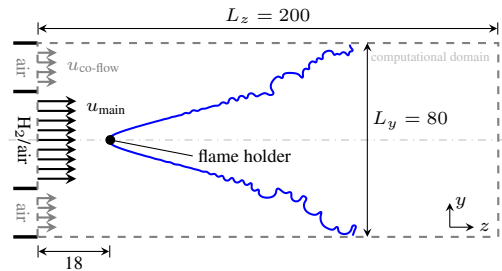
\begin{figure}[!htb]
    \centering
    \begin{tikzpicture} [scale=6.77/10.5]

        %symmetry line
        \draw[lightgray, dash dot] (-0.1,0)--++(10.2,0) node[right]{};

        %Burner schematic
        \draw[line width=1.3pt](0,-1)--++(0.5,0);
    	\draw[line width=1.3pt](0,1)--++(0.5,0);
        \draw[line width=1.3pt](0,-2)--++(0.5,0);
    	\draw[line width=1.3pt](0,2)--++(0.5,0);

        \node[rotate=-90] at (0.2,0) {\scriptsize H$_2$/air};
        \node[gray, rotate=-90] at (0.2,-1.5) {\scriptsize air};
        \node[gray, rotate=-90] at (0.2, 1.5) {\scriptsize air};

        % inflow
        \foreach \y in {-0.8,-0.6,-0.4,-0.2,0.0,0.2,0.4,0.6,0.8}
            \draw[vecteur] (0.5,\y) -- ++(0.8,0);
        
        % co-flow
        \foreach \y in {-1.8,-1.6,-1.4,-1.2,1.2,1.4,1.6,1.8}
            \draw[gray, vecteur] (0.5,\y) -- ++(0.4,0);

        \draw (1.92,0.65)node{\scriptsize $u_{\text{main}}$};
        \draw[gray] (1.66,1.5)node{\scriptsize $u_{\text{co-flow}}$};

        %flame front
        \begin{axis}[
            anchor=south west,
            at={(1.5cm,-2.37cm)},          
            width=7.65cm,             % scale down (pick any)
            axis equal image,
            hide axis,
            clip=false
            ]
            \addplot[
                blue,
                very thick,
                no markers
            ] table[
                col sep=comma,
                x=x,
                y=y
            ] {data/data_flame_front_simulation.csv};
        \end{axis}

        % Flame holder
    	\filldraw (2.0,0) circle (0.09);
        \draw[-] (2.0, 0)--++(1.7,-0.33) node[right]{\scriptsize flame holder};
        \draw[|<->|] (0.5,-2.3)--++(1.5,0) node[midway, below]{\scriptsize 18};

        % Coordinate system
        \draw[->] (9.,-1.8)--++(0,0.5) node[above, right]{\scriptsize $y$};
        \draw[->] (9.,-1.8)--++(0.5,0) node[above, right]{\scriptsize $z$};

        % Numerical domain
        \draw[gray, dashed, line width=0.8pt] (0.5,-2) rectangle (10,2);
        \node[lightgray, inner sep=2pt, anchor=center] at (8.6, 1.7) {\tiny computational domain};

        % lengths etc.
        \draw[|<->|] (0.5,2.3)--++(9.5,0) node[midway, above]{\scriptsize $L_{z}=200$};
        \draw[<->] (7.3,-2)--++(0,4) node[pos=0.6, right]{\scriptsize $L_{y}=80$};

    \end{tikzpicture}
    \vspace{1 pt}
   \caption{
    Schematic of the numerical setup.
    The dimensions of the computational domain are given in mm.
    Note that the figure is rotated by 90$^\circ$ relative to the experimental setup.
   }
  \label{fig:numerical_setup}
\addvspace{-0.1in}
\end{figure}

A two-dimensional numerical configuration in the $y$--$z$ plane is derived from the experimental setup.
It corresponds to a slice through the centre of the burner and can therefore be regarded as comparable to the OH-PLIF measurement plane.
A schematic of the numerical setup is shown in Fig.~\ref{fig:numerical_setup}.
As in the experiment, the circular flame holder (radius $1\,\mathrm{mm}$) is located $18\,\mathrm{mm}$ downstream of the inlet.
The computational domain has a length of $L_{z} = 200\,\mathrm{mm}$ and a width of $L_{y} = 80\,\mathrm{mm}$.

Non-reflecting boundary conditions are applied at the inlet and at the downstream and lateral outlets.
At the inlet, a uniform inflow profile with $u_{\mathrm{main}} = u_{\mathrm{bulk}}$ is imposed in the main stream region, $-20\,\mathrm{mm}\leq y \leq 20\,\mathrm{mm}$.
In the co-flow regions, $y < -20\,\mathrm{mm}$ and $y > 20\,\mathrm{mm}$, a uniform velocity profile with $u_{\mathrm{co-flow}}$ is prescribed, where $u_{\mathrm{co-flow}}$ is determined from the experimentally imposed co-flow volume flow rate for the corresponding equivalence ratio $\phi$.
The outlet pressure is fixed at $1\,\mathrm{atm}$.

The rod temperature is assumed to be $T_{\mathrm{rod}} = 900\,\mathrm{K}$.
A sensitivity study was conducted, maintaining constant equivalence ratio $\phi = 0.4$ and varying the rod temperature between $700\,\mathrm{K}$ and $1100\,\mathrm{K}$.
The best agreement with the experimental data, in terms of the flame anchoring position at the rod, was obtained for $T_{\mathrm{rod}} = 900\,\mathrm{K}$. This temperature is therefore used in all simulations.

For the simulations, the two limiting equivalence ratios considered in the experiments, $\phi=0.35$ and $\phi=0.40$, are selected, along with an intermediate equivalence ratio, $\phi=0.38$.

Simulations are performed with the finite-volume solver \textsc{CONVERGE}~5.1~\cite{Richards2026}, using second-order central differencing scheme for spatial discretization, a semi-implicit Crank-Nicolson scheme for time integration, and PISO pressure-velocity coupling.
Chemistry integration is performed with the SAGE solver employing the kinetic mechanism of Burke et al.~\cite{Burke2011}.
The mixture-averaged approximation is applied for species diffusion~\cite{Hirschfelder1964}, including an approximate model for Soret diffusion~\cite{Chapman1970}, as discussed, for example, in~\cite{Zirwes2025}.
A $200\,\mu$m Cartesian base grid is employed.
The flame is resolved using three levels of adaptive mesh refinement (AMR) to $25\,\mu$m,.
Therefore, the thermal flame thickness is resolved by at least 20 grid points in each case~\cite{Howarth2022}.

\subsection{Numerical Data Evaluation}

\begin{figure}
    \centering
    \includegraphics[width=\linewidth]{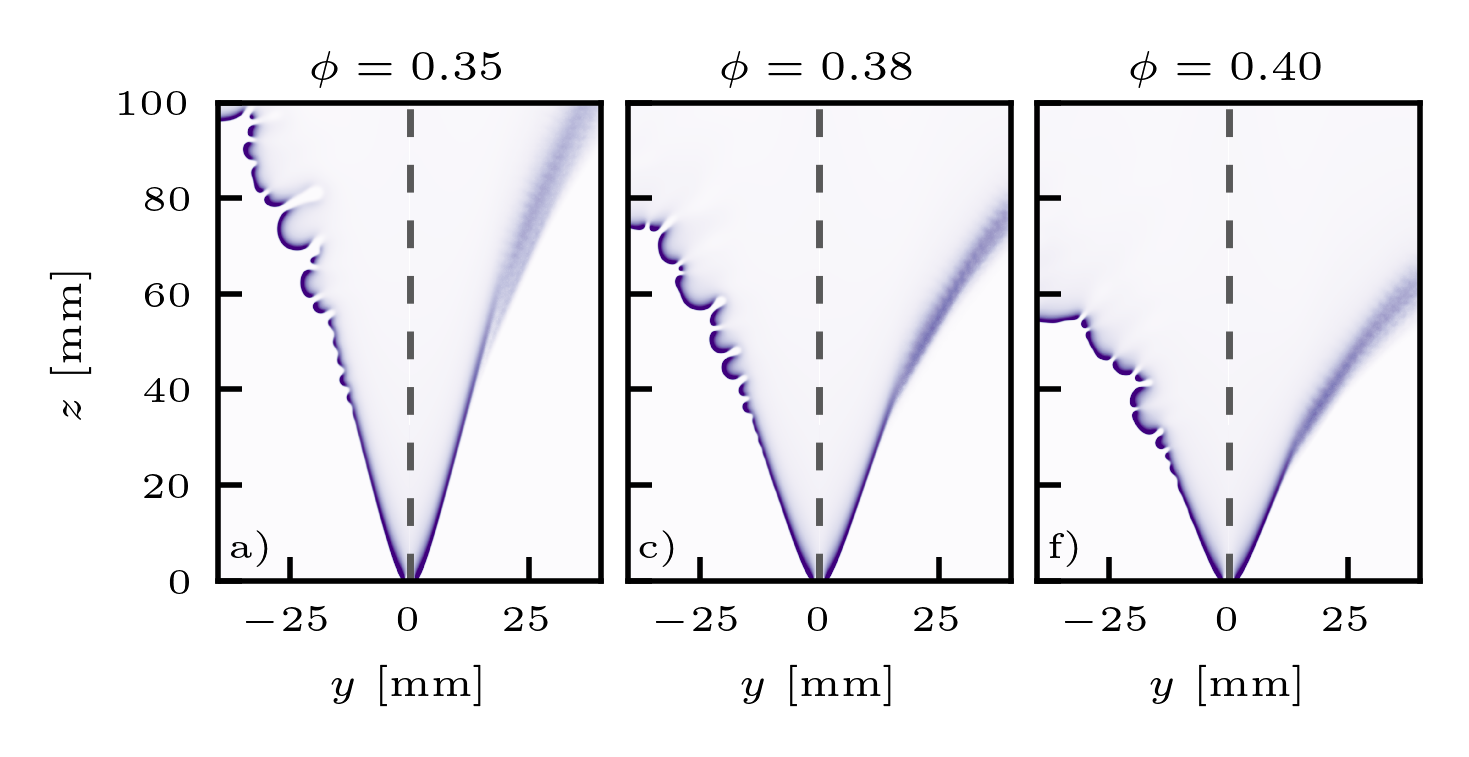}
    \caption{Selected instantaneous snapshots from the simulations (left panels) and the corresponding temporally averaged fields (right panels) of the recomputed OH-LIF signal for all simulated equivalence ratios.
    Note that the subfigure labels (a), (c), and (f) are chosen to be consistent with the experimental cases shown in Fig.~\ref{fig:AllOPs}.
    }
    \label{fig:OH_PLIF_simulation}
\end{figure}

Representative normalised OH-PLIF fields for the simulations, reconstructed from the thermochemical state of the numerical data following the approach detailed in~\cite{Popp2015}, are shown in Fig.~\ref{fig:OH_PLIF_simulation}.
As in Fig.~\ref{fig:AllOPs}, representative instantaneous fields are presented on the left, while the corresponding temporally averaged fields are shown on the right.

\begin{figure}
    \centering
    \includegraphics[width=\linewidth]{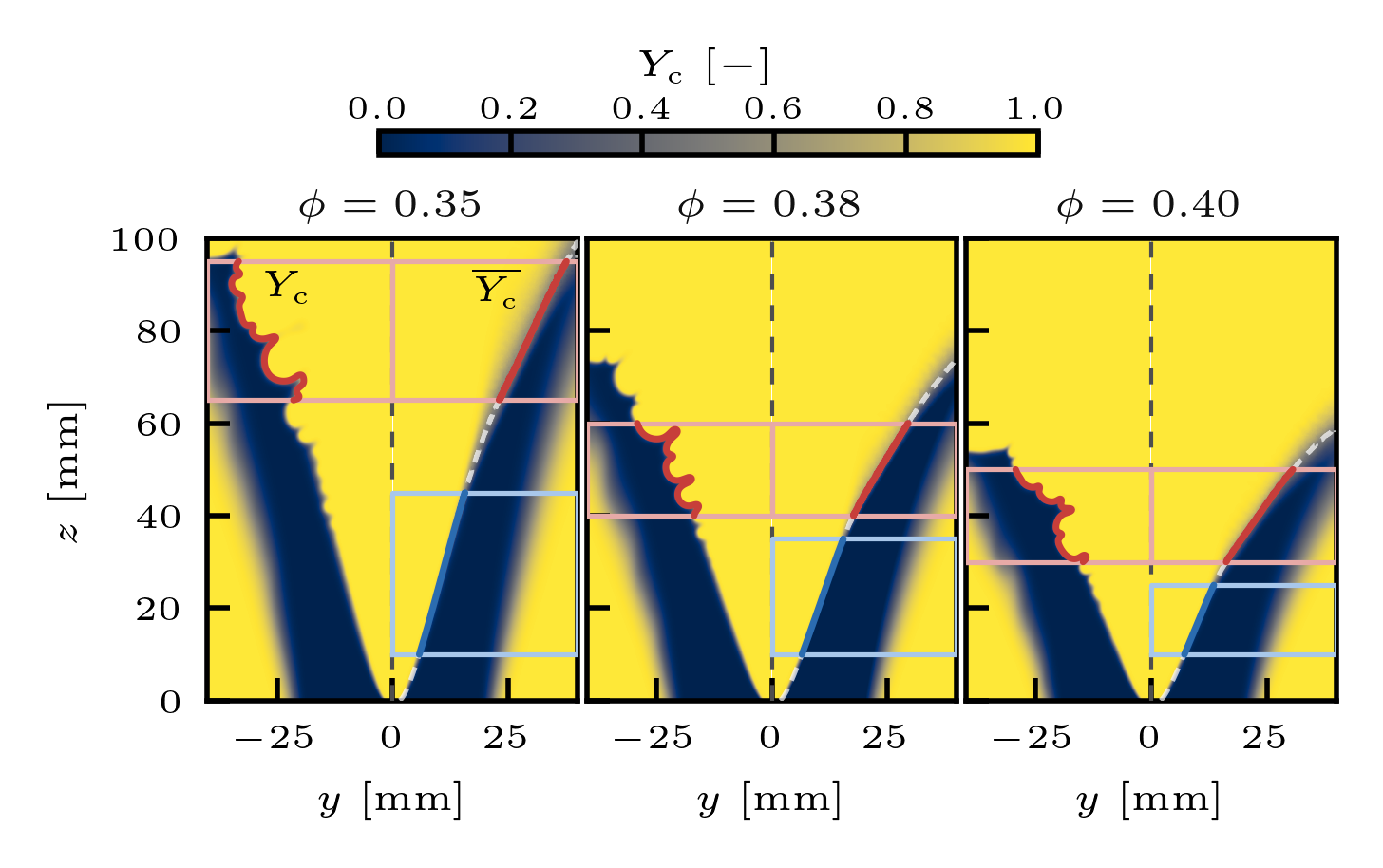}
    \caption{Selected instantaneous snapshots from the simulations (left panels) and the corresponding temporally averaged fields (right panels) of $Y_{\mathrm{c}}$ for all simulated equivalence ratios.
    In the right panels, the regions used to determine the stable branch (blue) and unstable branch (red) are highlighted.
    In addition, the corresponding linear fits to the mean flame front in these regions are indicated by blue and red lines, respectively.
    In the left panels, the flame front of the unstable branch is exemplarily highlighted for one instantaneous snapshot.}
    \label{fig:simulation_PV_analysis}
\end{figure}

An important advantage of the numerical simulations is that the full thermochemical state is available for the analysis.
Therefore, similar to previous studies in the literature, e.g.~\cite{Howarth2022}, the flame surface can be defined as an isosurface (or isoline) of the progress variable $Y_{\mathrm{c}} = 1 - {Y_{\ce{H2}}}/{Y_{\ce{H2},u}}$ and does not need to rely on OH measurements subject to uncertainty.
Based on this definition, the corresponding flame surfaces for the three simulations are shown in Fig.~\ref{fig:simulation_PV_analysis}, which also serves to illustrate the subsequent analysis procedure.
Note that, because the progress variable is defined in terms of the \ce{H2} mass fraction, it attains a value of unity in the co-flow region, where pure air is present.

Consistent with the experimental evaluation, separate regions are defined for the stable upstream branch and the unstable downstream branch in each simulation based on both instantaneous and averaged fields, as highlighted in Fig.~\ref{fig:simulation_PV_analysis}.
The region corresponding to the stable branch begins 10~mm downstream of the rod and extends, with an appropriate safety margin, to the location where the transition to the unstable branch occurs.
The region corresponding to the unstable branch begins, again with a safety margin from this transition point, and extends only as far as the flame remains sufficiently distant from the lateral outlet boundaries such that potential outlet effects can be excluded.

Within both regions, a linear fit is applied to the flame front in the averaged fields, where the mean flame front is defined by the $Y_{\mathrm{c}}=0.5$ isoline, consistent with Russell et al.~\cite{Russell2025}, who likewise used this value to define the mean flame front.
Based on these fits, the flame front angles in the stable and unstable regions are determined in a manner analogous to the experiment.
In addition, the linear fit in the unstable region is used to determine the mean flame surface area $A_0$ in that region.

For the instantaneous fields, the instantaneous flame surface area $A$ is determined over the same region used for the mean flame surface area and is subsequently averaged over all samples.
Here, also consistent with Russell et al.~\cite{Russell2025}, a value of $Y_{\mathrm{c}}=0.9$ is used to define the flame front.

Note that the co-flow is excluded from the analysis by retaining only the connected isolines with the largest $z$-values in each region.

For each case, the analysis is based on more than 500 snapshots sampled over a period of 200 flame times $\tau_L=\delta_L/S_L$, where $\delta_L$ denotes the thermal flame thickness of a corresponding one-dimensional reference flame at the same conditions.
Owing to the symmetry of the configuration, similarly to the analysis of the experimental measurements, both the left and right flame branches are available in each snapshot and are included in the analysis, effectively doubling the number of samples.

Additionally, the same processing method applied to the experimental data was also applied to the numerical simulation results, and the corresponding comparison is provided in the supplementary material. Very good agreement is observed between the two methods, further supporting the validity of the experimental methodology.

\section{Results and discussion \label{sec:results}} 
\begin{figure}[!h]
  \centering
  \includegraphics[width=1\columnwidth]{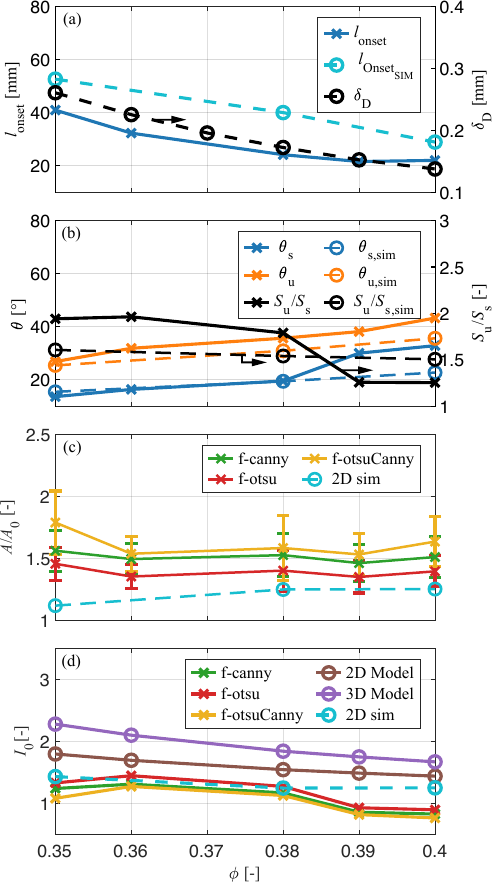}
  \captionsetup{aboveskip=4pt, belowskip=-10pt}
  \caption{Flame characteristics as a function of equivalence ratio $\phi$: (a) TDI onset location $l_{\mathrm{onset}}$ determined from the experiments and numerical simulations (left axis), as well as flame thickness $\delta_{\mathrm{D}}$ calculated using a multicomponent approach (right axis) (b) flame angles $\theta_{\mathrm{s}}$ and $\theta_{\mathrm{u}}$ (left axis) and the flame-speed ratio $S_{\mathrm{u}}/S_{\mathrm{s}}$ (right axis), for the experiments and numerical simulations; (c) area ratio $A/A_0$ and (d) stretch factor $I_0$, obtained from two experimental definitions and from the 2-D numerical simulations.
  In addition, two curves for $I_0$ computed from two models are shown, one based on 2-D (\cite[Eq.~22]{Howarth2022}) and the other on 3-D simulations (\cite[Eq.~7]{Howarth2023}) of TD unstable freely propagating \ce{H2} flames.}
  \label{fig:OnsetLengths}
\end{figure}

\addvspace{10pt}
For all equivalence ratios, two distinct flame branches and propagation regimes are identified in both the experimental measurements and the numerical simulations (see Fig.~\ref{fig:AllOPs}), with the distinction being most clearly visible in the instantaneous images (LHS).
In the upstream section, the flame front appears smooth and continuous (stable branch), whereas further downstream it transitions into a highly wrinkled structure characterised by pronounced cellular structures that are indicative of TDIs (unstable branch).
For $\phi = 0.37$, a distinct separation, or gap, in the OH-signal intensity between the stable and unstable branch was observed in the experiments; this behaviour is unique to this condition and the case is therefore excluded from further analysis.
The height of the TDI onset location decreases with increasing equivalence ratio in the experiments and the simulations (Figs.~\ref{fig:AllOPs} and \ref{fig:OH_PLIF_simulation}), showing a close correlation with the reduction in flame thickness $\delta_{\mathrm{D}}$ as depicted in Fig.~\ref{fig:OnsetLengths} (a).
Both the experimental and numerical datasets are evaluated using the same methodology as described in Sec.~\ref{sec:processing}.
Flame thicknesses are estimated using the diffusion thickness ($\delta_{\mathrm{D}}=\alpha/S_{\mathrm{L}}$), calculated using a mixture-averaged transport approach.
The mean images in Figs.~\ref{fig:AllOPs} and \ref{fig:OH_PLIF_simulation} (RHS) show a distinct change in flame inclination between the stable branch ($\theta_{\mathrm{s}}$) and TD-unstable one ($\theta_{\mathrm{u}}$).
For all equivalence ratios, the TD-unstable branch exhibits a steeper inclination, indicating that the TD-unstable flame front propagates more rapidly into the unburned mixture, i.e., it exhibits a higher effective flame consumption speed.
Assuming constant $u_{\mathrm{bulk}}$, the relation
\begin{equation}
    \frac{S_{\mathrm{u}}}{S_{\mathrm{s}}} = \frac{\sin(\theta_{\mathrm{u}})u_{\text{bulk}}}{\sin(\theta_{\mathrm{s}})u_{\text{bulk}}} = \frac{\sin(\theta_{\mathrm{u}})}{\sin(\theta_{\mathrm{s}})}
\end{equation} 
determines the ratio of the burning velocities of the two distinct flame branches for each operating condition.
The angles of both the stable and TD-unstable flame branches increase with equivalence ratio (Fig.~\ref{fig:OnsetLengths} (b)); however, the angle of the TD-unstable branch grows more slowly, resulting in a monotonic decrease in $S_{\mathrm{u}}/S_{\mathrm{s}}$.
An absolute value of $S_{\mathrm{s}}$ and $S_{\mathrm{u}}$ is not provided here, since neither the local flow field nor the strain rates were measured.
These will be addressed in future work using PIV measurements. For the present analysis, it is assumed that strain affects both branches in a similar manner, such that its influence largely cancels out in the relative comparison.

The occurrence of stronger TDIs at lower equivalence ratios is also predicted theoretically~\cite{Williams1985} and observed in earlier numerical studies~\cite{Berger2022b} and is also confirmed by the simulations conducted in the present study.
Enforcing unity-Lewis-number transport in the simulations suppresses the transition to the TD-unstable branch, confirming that the downstream flame wrinkling arises from TDI associated with the sub-unity Lewis number of lean \ce{H2} flames.

Despite the reduction in the normalised consumption speed $S_{\mathrm{u}}/S_{\mathrm{s}}$ with increasing equivalence ratio, the normalised flame surface area $A/A_0$, between the instantaneous and the averaged TD-unstable branch remains nearly constant, although a slight increase is observed in the numerical data, as shown in Fig.~\ref{fig:OnsetLengths} (c).
Three methodologies for the area determination from the experimental data are introduced, as detailed in the experimental data processing section, to quantify the uncertainties associated with defining the flame-surface area.
The different dependencies of ${S_{\mathrm{u}}}/{S_{\mathrm{s}}}$, decreasing with equivalence ratio, and of $A/A_0$, remaining approximately constant, represent the effect of a decrease of local reactivity that can be conveniently quantified by the stretch factor $I_0$, defined as \cite{Bray1991}:
\begin{equation}
I_0 = \frac{S_{\mathrm{u}}}{S_{\mathrm{s}}} \frac{A_0}{A} \, .
\end{equation}
In this analysis, it is implicitly assumed that similar strain rates act on the stable and unstable branches and that the ratio of their respective flame speeds, $S_{\mathrm{u}}/S_{\mathrm{s}}$, corresponds to the normalised flame consumption speed, $S_{\mathrm{C}}/S_{\mathrm{L}}$, i.e., the flame consumption speed of a TD-unstable flame normalised by the laminar 1-D flame speed.
The highest value of $I_0$ for the lowest equivalence ratio considered ($\phi = 0.35$) agrees well with the notion that TDIs are more pronounced for leaner mixtures, and that the positively-curved portions of the flame front are relatively more reactive, with locally faster flames due to localised enrichment and superadiabatic conditions, ultimately resulting in the acceleration of the global flame front.

A comparison between the numerical calculations and experimental results is shown in Fig.~\ref{fig:OnsetLengths} (d).
Additionally, results from two different empirical model fits (2-D and 3-D model) are shown.
These models were fitted using simulations of TD-unstable freely propagating flames in two~\cite{Howarth2022} and three~\cite{Howarth2023} dimensions, respectively, over a broad range of operating conditions.
They require only input parameters obtained from 1-D freely propagating flame simulations, primarily the instability parameter $\omega_{2}$ derived from linear stability analysis~\cite{Matalon2003}.
Both the 2-D~\cite{Howarth2022} and 3-D~\cite{Howarth2023} formulations of $I_0$ capture the same monotonic decrease in $I_0$ with increasing equivalence ratio.
The difference in magnitude between the $I_0$ values obtained from the experiments, numerically and by the models $I_0$ values is attributed primarily to variations in the definition of the flame front location (whether it is based on \ce{H2} consumption,  temperature,  OH distribution, or other criteria) as well as differences in the applied threshold levels.
In addition, discrepancies between the 2-D and 3-D models, as well as between the numerical simulations and the experiment, highlight the influence of 3-D effects.
These effects are only partially captured by the experimentally accessible 2-D measurement plane and are entirely absent in the 2-D simulations, indicating that future work should devise potential methodologies to obtain a ''3-D correction" of the experimental data.

\begin{figure}
    \centering
    \includegraphics[width=\linewidth]{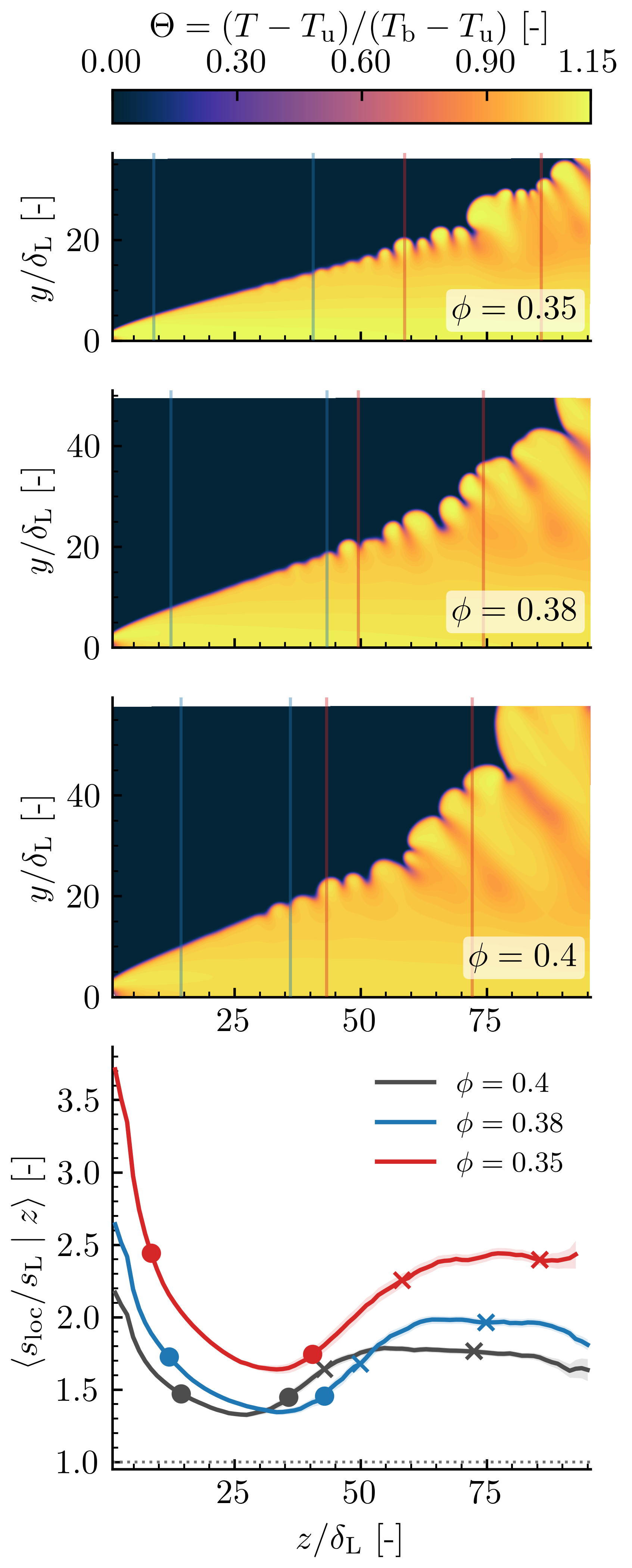}
    \caption{
        Top: Representative instantaneous normalised temperature $\Theta = (T-T_{\mathrm{u}})/(T_{\mathrm{b}}-T_{\mathrm{u}})$ fields of the upper flame branch for all three equivalence ratios $\phi$ from the numerical simulation.
        The axis are normalised by the 1-D thermal flame thickness $\delta_{\mathrm{L}}$.
        The blue and red lines highlight the regions, where the flame angles in the stable and unstable regions are determined, respectively.
        Note that the panels are rotated by 90$^\circ$ relative to the other figures.
        Bottom: Mean local flame speed $S_{\mathrm{loc}}/S_{\mathrm{L}}$ conditioned on the normalised streamwise coordinate $z/\delta_{\mathrm{L}}$ and averaged over all snapshots.
        Dots and crosses correspond to the regions highlighted in the top row.
    }
    \label{fig:sloc_fig_flame_length}
\end{figure}

In order to assess the validity of the proposed approach for estimating the reactivity factor $I_0$, the normalised mean local flame speed $S_{\mathrm{loc}}/S_{\mathrm{L}}$ is extracted from the numerical simulations and shown as a function of the streamwise location $z/\delta_{\mathrm{L}}$ in Fig.~\ref{fig:sloc_fig_flame_length} (bottom row) for the three equivalence ratios $\phi$ considered in the numerical simulations.
The methodology employs gradient trajectories of the progress variable to extract the local flame speed at a large number of locations along the flame front, following the approach outlined in \cite{Day2009, Howarth2022, Schneider2025a}.

The top row of Fig.~\ref{fig:sloc_fig_flame_length} shows exemplary normalised temperature fields over the normalised streamwise coordinate, scaled by the respective 1-D thermal flame thickness $\delta_{\mathrm{L}}$.
Several observations can be made.
As already indicated in the discussion of the onset length, the location of the transition from the stable to the unstable branch is approximately the same in normalised space for all cases.
At the flame holder, very high local flame speeds are obtained, which is attributed to the large curvature and strain effects in this region and the ignition process at the flame holder.
Downstream of this location, superadiabatic normalised temperatures with $\Theta>1$ emerge as a result of the presence of the flame holder.

Further downstream, a region is established in which the local flame speed remains approximately constant ($z/\delta_{\mathrm{L}} \approx 20\text{--}30$). This region is used to determine the stable branch of the flame and confirms the validity of the chosen extraction locations. Subsequently, as also visible in the instantaneous temperature snapshots, TDIs develop, and the normalised local flame speed increases significantly, coinciding with superadiabatic burning ($\Theta>1$). Here, another plateau region is established ($z/\delta_{\mathrm{L}}\gtrsim 50$), which agrees very well with the region from which the data for the unstable branch were extracted, thereby further confirming the validity of the chosen procedure.

At this point, it should also be noted that both regions appear to be slightly influenced by flame strain, increasing the local flame speed above the laminar flame speed in the stable region and above the freely-propagating flame speed of a freely-propagating TD-unstable flame in the unstable region.
Nevertheless, it is assumed that the proposed model yields a valid estimate of $I_0$, since strain affects both regions and therefore at least partially cancels out in the estimation of $I_0$ based on the different flame angles. In future work, this aspect will be investigated in more detail, for example by incorporating particle image velocimetry (PIV) measurements in the experiments.

\section{Conclusion\label{sec:conclusion}} \addvspace{10pt}

A novel experimental methodology for determining the stretch factor $I_0$ has been developed and demonstrated.
The method is tailored to V-shaped flames stabilised on a cylindrical flame holder and requires only instantaneous and temporally averaged snapshots of a planar (2-D) flame diagnostic, such as OH-PLIF.
In such a configuration, as shown in the present study by both experiments and 2-D numerical simulations, a flat and stable branch forms downstream of the flame holder and, beyond a certain distance from the flame holder, transitions into a branch dominated by thermodiffusive instabilities, as evidenced by the development of cellular flame structures.

Owing to the locally enhanced reactivity associated with these instabilities, i.e.\ an increased $I_0$, and the increased instantaneous flame wrinkling relative to the mean flame brush, the flame consumption speed, $S_{\mathrm{c}}$, is also increased in this branch.
This manifests itself in a larger inclination angle of the TD-unstable branch relative to the streamwise direction compared to the stable branch.
The methodology exploits this behaviour by determining the flame-speed ratio between the unstable and stable branch $S_{\mathrm{u}}/S_{\mathrm{s}}$ geometrically from the corresponding flame angles.
As a first approximation, the flame speed in the stable branch is assumed to be equal to the 1-D laminar flame speed, such that $S_{\mathrm{u}}/S_{\mathrm{s}} \approx S_{\mathrm{c}}/S_{\mathrm{L}}$.

In addition, the mean flame front area $A_0$ and the temporally averaged instantaneous flame front area $A$ are determined in the unstable branch from the averaged and instantaneous fields, respectively.
Based on the relationship between $S_{\mathrm{c}}/S_{\mathrm{L}}$ and $A/A_0$, the stretch factor is then obtained as
$I_0 = {S_{\mathrm{c}}}/{S_{\mathrm{L}}} \cdot {A_0}/{A} \approx  {S_{\mathrm{u}}}/{S_{\mathrm{s}}} \cdot {A_0}/{A}$.

The analysis of both experiments and simulations reveals that the inclination difference between the upstream stable branch and the downstream TD-unstable branch decreases systematically with increasing equivalence ratio, corresponding to a reduced normalised flame speed.
In contrast, the corresponding area ratio remains nearly constant over all investigated conditions.
As a result, $I_0$ decreases with increasing equivalence ratio.
This trend is consistent with theoretical expectations and numerical models, indicating consistent underlying physical behaviour.

The remaining discrepancies between the experimental, numerical, and model-based results are attributed to differences and residual uncertainties in the estimation of the flame-surface area, as well as to the influence of local strain and 3-D effects, such as an enhanced thermodiffusive response and, consequently, a more strongly curved flame front.
While such 3-D effects are inherently present in the experiments, they are not captured in the 2-D OH-PLIF measurement plane.
Furthermore, these are not represented at all in the 2-D numerical simulations.
Future work should therefore investigate the influence of strain and 3-D effects, as well the transition from the stable to the unstable branch, and improve the procedure to determine the flame-surface area.

\acknowledgement{CRediT authorship contribution statement} \addvspace{10pt}

\textbf{MM}: Investigation, Data curation, Software, Visualisation, Writing- Orig. draft, Review \& Editing;
\textbf{CM}: Investigation, Data curation, Software, Visualisation, Writing- Orig. draft, Review \& Editing;
\textbf{ARWM}: Conceptualisation, Writing- Orig. draft, Review \& Editing;
\textbf{MS}: Conceptualization, Investigation, Data curation, Software, Visualisation, Writing- Orig. draft, Review \& Editing;
\textbf{BT}: Conceptualization, Investigation, Data curation, Software, Visualisation, Writing- Orig. draft, Review \& Editing;
\textbf{CH}: Funding acquisition, Supervision, Writing - Review \& Editing;
\textbf{AG}: Conceptualisation, Methodology, Writing- Review \& Editing;
\textbf{AD}: Conceptualization, Funding acquisition, Supervision, Writing- Review \& Editing.

\acknowledgement{Declaration of competing interest} \addvspace{10pt}
The authors declare that they have no known competing financial interests or personal relationships that could have appeared to influence the work reported in this paper.

\acknowledgement{Acknowledgements} \addvspace{10pt}
This research has been funded by the German Research Foundation (DFG) in the framework of SPP 2419 HyCAM (Project number 523792378).
\textsc{CONVERGE} licenses and technical support were provided by Convergent Science.

% -------------------------------------------------------------------- %
% -------------------------------------------------------------------- %
% -------------------------------------------------------------------- %
\footnotesize
\baselineskip 9pt

% -------------------------------------------------------------------- %
% -------------------------------------------------------------------- %
% -------------------------------------------------------------------- %
\clearpage
\thispagestyle{empty}
\bibliographystyle{proci}
\bibliography{PROCI_LaTeX}

@article{canny2009computational,
  title = {{A Computational Approach to Edge Detection}},
  volume = {PAMI-8},
  DOI = {10.1109/tpami.1986.4767851},
  number = {6},
  journal = {IEEE Trans. Pattern Anal. Mach. Intell.},
  publisher = {Institute of Electrical and Electronics Engineers (IEEE)},
  author = {Canny,  John},
  year = {1986},
  month = nov,
  pages = {679–698}
}

@article{otsu1979threshold,
  title = {{A Threshold Selection Method from Gray-Level Histograms}},
  volume = {9},
  DOI = {10.1109/tsmc.1979.4310076},
  number = {1},
  journal = {IEEE Trans. Syst. Man Cybern.},
  publisher = {Institute of Electrical and Electronics Engineers (IEEE)},
  author = {Otsu,  Nobuyuki},
  year = {1979},
  month = jan,
  pages = {62–66}
}

@article{Nicolai2025,
  title = {Laminar and turbulent hydrogen-enriched methane flames: Interaction of thermodiffusive instabilities and local fuel demixing},
  volume = {41},
  DOI = {10.1016/j.proci.2025.105885},
  journal = {Proc. Combust. Inst.},
  publisher = {Elsevier BV},
  author = {Nicolai,  Hendrik and Schuh,  Vinzenz and B\"{a}hr,  Antonia and Schneider,  Max and Rong,  Felix and Kaddar,  Driss and Bode,  Mathis and Hasse,  Christian},
  year = {2025},
  pages = {105885}
}

@article{Schneider2025a,
  title = {{Flame-wall interaction of thermodiffusively unstable hydrogen/air flames, Part I: Characterization of governing physical phenomena}},
  volume = {279},
  DOI = {10.1016/j.combustflame.2025.114320},
  journal = {Combust. Flame},
  publisher = {Elsevier BV},
  author = {Schneider,  Max and Nicolai,  Hendrik and Schuh,  Vinzenz and Steinhausen,  Matthias and Hasse,  Christian},
  year = {2025},
  pages = {114320}
}

@article{Schneider2025b,
  title = {{Flame-wall interaction of thermodiffusively unstable hydrogen/air flames, Part II: Parametric variations of equivalence ratio, temperature, and pressure}},
  volume = {279},
  DOI = {10.1016/j.combustflame.2025.114319},
  journal = {Combust. Flame},
  publisher = {Elsevier BV},
  author = {Schneider,  Max and Nicolai,  Hendrik and Schuh,  Vinzenz and Steinhausen,  Matthias and Hasse,  Christian},
  year = {2025}
}

@article{Schneider2025c,
  title = {Combustion modelling for the flame–wall interaction of thermodiffusively unstable hydrogen/air flames},
  volume = {1029},
  DOI = {10.1017/jfm.2026.11200},
  journal = {J. Fluid Mech.},
  publisher = {Cambridge University Press (CUP)},
  author = {Schneider,  Max and Rong,  Felix Zijie and Hasse,  Christian and Nicolai,  Hendrik},
  year = {2026},
  month = feb 
}

@article{Dreizler2021,
  title = {The role of combustion science and technology in low and zero impact energy transformation processes},
  volume = {7},
  DOI = {10.1016/j.jaecs.2021.100040},
  journal = {Appl. Energy Combust. Sci.},
  publisher = {Elsevier BV},
  author = {Dreizler,  A. and Pitsch,  H. and Scherer,  V. and Schulz,  C. and Janicka,  J.},
  year = {2021},
  month = sep,
  pages = {100040}
}

@article{Howarth2022,
  title = {An empirical characteristic scaling model for freely-propagating lean premixed hydrogen flames},
  volume = {237},
  DOI = {10.1016/j.combustflame.2021.111805},
  journal = {Combust. Flame},
  publisher = {Elsevier BV},
  author = {Howarth,  T.L. and Aspden,  A.J.},
  year = {2022},
  pages = {111805}
}

@article{Howarth2023,
  title={{Thermodiffusively-unstable lean premixed hydrogen flames: Phenomenology, empirical modelling, and thermal leading points}},
  volume={253},
  DOI={10.1016/j.combustflame.2023.112811},
  journal={Combust. Flame},
  publisher={Elsevier BV},
  author={Howarth, T. and Hunt, E. and Aspden, A.},
  year={2023},
  pages={112811} ,
}

@article{Berger2022c,
  title = {Synergistic interactions of thermodiffusive instabilities and turbulence in lean hydrogen flames},
  volume = {244},
  DOI = {10.1016/j.combustflame.2022.112254},
  journal = {Combust. Flame},
  publisher = {Elsevier BV},
  author = {Berger,  Lukas and Attili,  Antonio and Pitsch,  Heinz},
  year = {2022},
  pages = {112254}
}

@article{Berger2022b,
  title = {{Intrinsic instabilities in premixed hydrogen flames: parametric variation of pressure, equivalence ratio, and temperature. Part 2 – Non‐linear regime and flame speed enhancement}},
  volume = {240},
  DOI = {10.1016/j.combustflame.2021.111936},
  journal = {Combust. Flame},
  publisher = {Elsevier BV},
  author = {Berger,  Lukas and Attili,  Antonio and Pitsch,  Heinz},
  year = {2022},
  pages = {111936}
}

@article{Altantzis2012,
  title = {Hydrodynamic and thermodiffusive instability effects on the evolution of laminar planar lean premixed hydrogen flames},
  volume = {700},
  DOI = {10.1017/jfm.2012.136},
  journal = {J. Fluid Mech.},
  publisher = {Cambridge University Press},
  author = {Altantzis,  C. and Frouzakis,  C. E. and Tomboulides,  A. G. and Matalon,  M. and Boulouchos,  K.},
  year = {2012},
  pages = {329–361}
}

@article{Kadowaki2005b,
  title = {The unstable behavior of cellular premixed flames induced by intrinsic instability},
  volume = {30},
  DOI = {10.1016/j.proci.2004.07.041},
  number = {1},
  journal = {Proc. Combust. Inst.},
  publisher = {Elsevier BV},
  author = {Kadowaki,  Satoshi and Suzuki,  Hiroshi and Kobayashi,  Hideaki},
  year = {2005},
  pages = {169–176}
}

@article{Aspden2015,
  title = {Turbulence-chemistry interaction in lean premixed hydrogen combustion},
  volume = {35},
  DOI = {10.1016/j.proci.2014.08.012},
  number = {2},
  journal = {Proc. Combust. Inst.},
  publisher = {Elsevier BV},
  author = {Aspden,  A.J. and Day,  M.S. and Bell,  J.B.},
  year = {2015},
  pages = {1321–1329}
}

@article{Pitsch2024,
  title = {The transition to sustainable combustion: {Hydrogen}- and carbon-based future fuels and methods for dealing with their challenges},
  volume = {40},
  DOI = {10.1016/j.proci.2024.105638},
  number = {1–4},
  journal = {Proc. Combust. Inst.},
  publisher = {Elsevier BV},
  author = {Pitsch,  Heinz},
  year = {2024},
  pages = {105638}
}

@article{Kosaka2018,
  title = {Wall heat fluxes and {CO} formation/oxidation during laminar and turbulent side-wall quenching of methane and DME flames},
  volume = {70},
  DOI = {10.1016/j.ijheatfluidflow.2018.01.009},
  journal = {Int. J. Heat Fluid Flow},
  publisher = {Elsevier BV},
  author = {Kosaka,  Hidemasa and Zentgraf,  Florian and Scholtissek,  Arne and Bischoff,  Lothar and H\"{a}ber,  Thomas and Suntz,  Rainer and Albert,  Barbara and Hasse,  Christian and Dreizler,  Andreas},
  year = {2018},
  month = apr,
  pages = {181–192}
}

@article{Bradley1994,
  title = {The development of instabilities in laminar explosion flames},
  volume = {99},
  DOI = {10.1016/0010-2180(94)90049-3},
  number = {3–4},
  journal = {Combust. Flame},
  publisher = {Elsevier BV},
  author = {Bradley,  Derek and Harper,  C.M.},
  year = {1994},
  pages = {562–572}
}

@article{Li2025,
  title = {{Experimental Insights into Thermodiffusive Instabilities in Lean Hydrogen Combustion}},
  DOI = {10.1016/j.eng.2025.09.016},
  journal = {Eng.},
  publisher = {Elsevier BV},
  author = {Li,  Tao and B\"{o}hm,  Benjamin and Dreizler,  Andreas},
  year = {2025},
}

@article{Shi2024a,
  title = {{Structures of Laminar Lean Premixed H2/CH4/Air Polyhedral Flames: Effects of Flow Velocity,  H2 Content and Equivalence Ratio}},
  volume = {113},
  DOI = {10.1007/s10494-024-00561-3},
  number = {4},
  journal = {Flow Turbul. Combust.},
  publisher = {Springer Science and Business Media LLC},
  author = {Shi,  Shuguo and Breicher,  Adrian and Schultheis,  Robin and Hartl,  Sandra and Barlow,  Robert S. and Geyer,  Dirk and Dreizler,  Andreas},
  year = {2024},
  pages = {1081–1110}
}

@article{Chaib2024,
title = {An experimental marker of thermo-diffusive instability in hydrogen-enriched flames},
journal = {Proc. Combust. Inst.},
volume = {40},
number = {1},
pages = {105763},
year = {2024},
issn = {1540-7489},
doi = {https://doi.org/10.1016/j.proci.2024.105763},
author = {Oussama Chaib and Simone Hochgreb and Isaac Boxx},
}

@book{Zeldovich1944,
  title={Theory of combustion and detonation of gases},
  author={Zeldovich, YB},
  year={1944},
  publisher={Acad. Sci. USSR.}
}

@article{Markstein1949,
  title={Cell structure of propane flames burning in tubes},
  author={Markstein, GH},
  journal={J. Chem. Phys.},
  volume={17},
  number={4},
  pages={428--429},
  year={1949},
  publisher={American Institute of Physics}
}

@book{Williams1985,
	address = {Menlo Park, Calif},
	edition = {2nd ed},
	series = {Combustion science and engineering series},
	title = {Combustion theory: the fundamental theory of chemically reacting flow systems},
	isbn = {978-0-8053-9801-4},
	shorttitle = {Combustion theory},
	publisher = {Benjamin/Cummings Pub. Co},
	author = {Williams, F. A.},
	year = {1985},
}

@article{Aspden2011,
	title = {Turbulence–flame interactions in lean premixed hydrogen: transition to the distributed burning regime},
	volume = {680},
	shorttitle = {Turbulence–flame interactions in lean premixed hydrogen},
	url = {https://www.cambridge.org/core/product/identifier/S0022112011001649/type/journal_article},
	doi = {10.1017/jfm.2011.164},
	journal = {J. Fluid Mech.},
	author = {Aspden, A. J. and Day, M. S. and Bell, J. B.},
	year = {2011},
	pages = {287--320}
}

@article{Aspden2019,
	title = {Towards the distributed burning regime in turbulent premixed flames},
	volume = {871},
	urldate = {2024-12-27},
	journal = {J. Fluid Mech.},
	author = {Aspden, A. J. and Day, M. S. and Bell, J. B.},
	year = {2019},
	pages = {1--21}
}

@article{Aspden2024,
	title = {Three-dimensional phenomenology of freely-propagating thermodiffusively-unstable lean premixed hydrogen flames},
	volume = {40},
	doi = {10.1016/j.proci.2024.105634},
	language = {en},
	number = {1-4},
	urldate = {2024-08-28},
	journal = {Proc. Combust. Inst.},
	author = {Aspden, A.J. and Howarth, T.L. and Hunt, E.F.},
	year = {2024},
	pages = {105634}
}

@article{Berger2024,
	title = {Effects of {Karlovitz} number variations on thermodiffusive instabilities in lean turbulent hydrogen jet flames},
	volume = {40},
	doi = {10.1016/j.proci.2024.105219},
	number = {1-4},
	urldate = {2024-08-16},
	journal = {Proc. Combust. Inst.},
	author = {Berger, L. and Attili, A. and Gauding, M. and Pitsch, H.},
	year = {2024},
	pages = {105219}
}

@article{Matalon2003,
  title = {Hydrodynamic theory of premixed flames: effects of stoichiometry,  variable transport coefficients and arbitrary reaction orders},
  volume = {487},
  DOI = {10.1017/s0022112003004683},
  journal = {J. Fluid Mech.},
  publisher = {Cambridge University Press (CUP)},
  author = {Matalon,  M. and Cui,  C. and Bechtold,  J. K.},
  year = {2003},
  month = jun,
  pages = {179–210}
}

@article{Sivashinsky1977,
  title = {{Diffusional-Thermal Theory of Cellular Flames}},
  volume = {15},
  url = {http://dx.doi.org/10.1080/00102207708946779},
  DOI = {10.1080/00102207708946779},
  number = {3–4},
  journal = {Combust. Sci. Technol.},
  publisher = {Informa UK Limited},
  author = {Sivashinsky,  G. I.},
  year = {1977},
  month = jan,
  pages = {137–145}
}

@article{Shi2024b,
  title = {{Internal flame structures of thermo-diffusive lean premixed H2/air flames with increasing turbulence}},
  volume = {40},
  url = {http://dx.doi.org/10.1016/j.proci.2024.105225},
  DOI = {10.1016/j.proci.2024.105225},
  number = {1–4},
  journal = {Proc. Combust. Inst.},
  publisher = {Elsevier BV},
  author = {Shi,  Shuguo and Schultheis,  Robin and Barlow,  Robert S. and Geyer,  Dirk and Dreizler,  Andreas and Li,  Tao},
  year = {2024},
  pages = {105225}
}

@article{Porath2025,
  title = {Low velocity streaks combined with intrinsic flame instabilities provoke boundary layer flashback in a turbulent premixed jet-stabilized hydrogen flame},
  volume = {278},
  url = {http://dx.doi.org/10.1016/j.combustflame.2025.114236},
  DOI = {10.1016/j.combustflame.2025.114236},
  journal = {Combust. Flame},
  publisher = {Elsevier BV},
  author = {Porath,  P. and Altenburg,  L.A. and Klein,  S.A. and Tummers,  M.J. and Ghani,  A.},
  year = {2025},
  month = aug,
  pages = {114236}
}

@article{Marburger2026,
  title = {Comparative experimental study of flame–wall interaction for hydrogen and methane},
  DOI = {10.1016/j.jaecs.2026.100485},
  journal = {Appl. Energy Combust. Sci.},
  publisher = {Elsevier BV},
  author = {Marburger,  Marcel and M\"{o}ller,  Christoph and Schneider,  Max and Macfarlane,  Andrew and Dreizler,  Andreas},
  year = {2026},
  month = mar,
  pages = {100485}
}

@article{Ye2025,
  title = {Analysis of thermodiffusive instabilities and flame front wrinkling in a hydrogen-fueled engine},
  volume = {41},
  DOI = {10.1016/j.proci.2025.105884},
  journal = {Proc. Combust. Inst.},
  publisher = {Elsevier BV},
  author = {Ye,  Pedro and Erhard,  Jannick and Welch,  Cooper and Shi,  Hao and Dreizler,  Andreas and B\"{o}hm,  Benjamin},
  year = {2025},
  pages = {105884}
}

@article{Traut2026,
  title = {Numerical modeling of lean hydrogen spark-ignition engines: On the role of intrinsic instabilities},
  volume = {26},
  DOI = {10.1016/j.jaecs.2026.100473},
  journal = {Appl. Energy Combust. Sci.},
  publisher = {Elsevier BV},
  author = {Traut,  Benjamin and Schuh,  Vinzenz and Hasenzahl,  Max and Kircher,  Magnus and Nicolai,  Hendrik and Hasse,  Christian},
  year = {2026},
  month = jun,
  pages = {100473}
}

@article{Rieth2023,
  title = {The effect of pressure on lean premixed hydrogen-air flames},
  volume = {250},
  DOI = {10.1016/j.combustflame.2022.112514},
  journal = {Combust. Flame},
  publisher = {Elsevier BV},
  author = {Rieth,  Martin and Gruber,  Andrea and Chen,  Jacqueline H.},
  year = {2023},
  month = apr,
  pages = {112514}
}

@article{Day2009,
  title = {Turbulence effects on cellular burning structures in lean premixed hydrogen flames},
  volume = {156},
  DOI = {10.1016/j.combustflame.2008.10.029},
  number = {5},
  journal = {Combust. Flame},
  publisher = {Elsevier BV},
  author = {Day,  Marc and Bell,  John and Bremer,  Peer-Timo and Pascucci,  Valerio and Beckner,  Vince and Lijewski,  Michael},
  year = {2009},
  month = may,
  pages = {1035–1045}
}

@manual{Richards2026,
  author    = {Richards, K. J. and Senecal, P. K. and Pomraning, E.},
  title     = {{CONVERGE 5.1}},
  organization = {Convergent Science},
  year      = {2026},
}

@article{Burke2011,
  title = {{Comprehensive H2/O2 kinetic model for high‐pressure combustion}},
  volume = {44},
  ISSN = {1097-4601},
  DOI = {10.1002/kin.20603},
  number = {7},
  journal = {Int. J. Chem. Kinet.},
  publisher = {Wiley},
  author = {Burke,  Michael P. and Chaos,  Marcos and Ju,  Yiguang and Dryer,  Frederick L. and Klippenstein,  Stephen J.},
  year = {2011},
  pages = {444–474}
}

@book{Hirschfelder1964,
  title={The molecular theory of gases and liquids},
  author={Hirschfelder, Joseph O and Curtiss, Charles F and Bird, R Byron},
  year={1964},
  publisher={John Wiley \& Sons}
}

@book{Chapman1970,
  author    = {Chapman, S. and Cowling, T. G.},
  title     = {The Mathematical Theory of Non-Uniform Gases},
  publisher = {Cambridge University Press},
  address   = {Cambridge},
  year      = {1970}
}

@article{Zirwes2025,
  title = {{Assessment of Approximate Soret Diffusion Models for Hydrogen and Ammonia Combustion}},
  volume = {115},
  DOI = {10.1007/s10494-025-00680-5},
  number = {4},
  journal = {Flow  Turb. Combust.},
  publisher = {Springer Science and Business Media LLC},
  author = {Zirwes,  Thorsten and Kronenburg,  Andreas},
  year = {2025},
  month = jul,
  pages = {1631–1650}
}

@article{Popp2015,
  title = {{LES flamelet-progress variable modeling and measurements of a turbulent partially-premixed dimethyl ether jet flame}},
  volume = {162},
  url = {http://dx.doi.org/10.1016/j.combustflame.2015.05.004},
  DOI = {10.1016/j.combustflame.2015.05.004},
  number = {8},
  journal = {Combust. Flame},
  publisher = {Elsevier BV},
  author = {Popp,  Sebastian and Hunger,  Franziska and Hartl,  Sandra and Messig,  Danny and Coriton,  Bruno and Frank,  Jonathan H. and Fuest,  Frederik and Hasse,  Christian},
  year = {2015},
  month = aug,
  pages = {3016–3029}
}

@article{Russell2025,
  title = {{Turbulence-flame interactions in high-Karlovitz-number lean premixed hydrogen piloted jet flames}},
  volume = {41},
  DOI = {10.1016/j.proci.2025.105868},
  journal = {Proc. Combust. Inst.},
  publisher = {Elsevier BV},
  author = {Russell,  G.S. and Howarth,  T.L. and Skiba,  A.W. and Carter,  C.D. and Aspden,  A.J.},
  year = {2025},
  pages = {105868}
}

@article{Zentgraf2022,
title = {Detailed assessment of the thermochemistry in a side-wall quenching burner by simultaneous quantitative measurement of {CO$_2$}, {CO} and temperature using laser diagnostics},
  volume = {235},
  DOI = {10.1016/j.combustflame.2021.111707},
  journal = {Combust. Flame},
  publisher = {Elsevier BV},
  author = {Zentgraf,  Florian and Johe,  Pascal and Steinhausen,  Matthias and Hasse,  Christian and Greifenstein,  Max and Cutler,  Andrew D. and Barlow,  Robert S. and Dreizler,  Andreas},
  year = {2022},
  month = jan,
  pages = {111707}
}

@article{Bray1991,
    title = {Some applications of {Kolmogorov}’s turbulence research in the field of combustion},
    volume = {434},
    DOI = {10.1098/rspa.1991.0090},
    number = {1890},
    journal = {Proc. R. Soc. Lond. A},
    publisher = {The Royal Society},
    author = {Bray, Kenneth Noel Corbett and Cant, R. S.},
    editor = {Hunt, Julian Charles Roland and Phillips, Owen Martin and Williams, David},
    year = {1991},
    month = jul,
    pages = {217–240} }

@article{Yang2018,
  title={Extreme role of preferential diffusion in turbulent flame propagation},
  author={Yang, Sheng and Saha, Abhishek and Liang, Wenkai and Wu, Fujia and Law, Chung K},
  journal={Combust. Flame},
  volume={188},
  pages={498--504},
  year={2018},
  publisher={Elsevier}
}

% -------------------------------------------------------------------- %
% -------------------------------------------------------------------- %
% -------------------------------------------------------------------- %

\newpage

\small
\baselineskip 10pt

% -------------------------------------------------------------------- %
% -------------------------------------------------------------------- %
% -------------------------------------------------------------------- %

\end{document}